\documentclass[fleqn,10pt]{wlscirep}
\usepackage[utf8]{inputenc}
\usepackage[T1]{fontenc}

\title{Surface reconstruction limited magnetism of the nodal loop semimetal Ca$_3$P$_2$}

\author[1,*]{Assem Alassaf}
\author[2]{János Koltai}
\author[3,4]{Amador García-Fuente}
\author[1,5]{László Oroszlány}

\affil[1]{
 Department of Physics of Complex Systems, ELTE E{\" o}tv{\" o}s Loránd University, P{\'a}zmány P{\'e}ter s{\' e}t{\' a}ny 1/A, 1117 Budapest, Hungary}
\affil[2]{
 Department of Biological Physics, ELTE E{\" o}tv{\" o}s Loránd University, P{\'a}zmány P{\'e}ter s{\' e}t{\' a}ny 1/A, 1117 Budapest, Hungary
}
\affil[3]{Departamento de Física,  Universidad de Oviedo,  33007 Oviedo, Spain}
\affil[4]{Centro de Investigación en Nanomateriales y Nanotecnología, Universidad de Oviedo-CSIC, 33940, El Entrego, Spain}
\affil[5]{Wigner Research Centre for Physics, Konkoly-Thege M. út 29-33, H-1121 Budapest, Hungary}%
\affil[*]{assem.al.assaf.abd.alrham@ttk.elte.hu}

\keywords{Nodal loop semimetals, Surface magnetism, Ab initio}

\begin{abstract}
Nodal loop semimetals are topological materials with drumhead surface states characterized by reduced kinetic energy. If the Fermi energy of such a system is near these nondispersive states interaction among charge carriers substantially impacts their electronic structure. The emergence of magnetism in these surface states is one of the possible consequences. Ca$_3$P$_2$ an already synthesized material possesses a remarkably large nodal loop which is situated exactly at the Fermi level of the bulk system. In the present work, we investigate how surface magnetism is impacted by surface reconstruction and lattice termination in finite slabs in this material.
We show that a slight deviation from the stoichiometric occupation of Ca sites results in the stabilization of magnetic patterns.
\end{abstract}
\begin{document}

\flushbottom
\maketitle

\thispagestyle{empty}

\section*{Introduction}

In nodal loop semimetals, band degeneracies form a closed loop in the three-dimensional Brillouin zone.\cite{NODAL_REV_yang2018symmetry} In these topological materials, topological protection is linked to additional discrete symmetries of the crystal.\cite{fang2015topological} \emph{Ab initio} calculations paved the way in the search for the physical realization of nodal loop semimetals. \cite{NODAL_first_PhysRevLett.115.036807} In many of these calculations, it was highlighted, that spin-orbit coupling in most cases splits up the nodal structure into discrete Weyl points. 
Compelling evidence for the existence of a nodal loop in ZrSiS was found in quantum oscillation measurements. \cite{matusiak2017thermoelectric} The experimental isolation of the stoichiometric compound Ca$_3$P$_2$ is also considered an important step towards the technological exploitation of nodal semimetals. \cite{ca3p2_experiment} Spin-orbit coupling is expected to be negligible in this system, which is composed of relatively lightweight elements. Furthermore, first-principles calculations of the bulk spectrum revealed a large nodal loop precisely at the Fermi level.
Drumhead states on the surfaces of nodal semimetals, are defined by the surface projection of the nodal structure of the bulk. These topological boundary states have small kinetic energy and are thus susceptible to interactions, making them an ideal platform for emergent surface magnetism. \cite{Roy_Interacting_nodal_PhysRevB.96.041113} Spin-polarized surface states in SnTaS$_2$ samples were for instance recently reported.\cite{RECENT_NODAL_ARPES_PhysRevB.107.045142} Interaction-induced magnetic properties have already been studied theoretically \cite{Campetella_PhysRevB.101.165437,ED_paper2021exchange} and observed experimentally \cite{zhou2021half,hagymasi2022observation} in rhombohedral graphite, a material also exhibiting a helical nodal line in its bulk. In two dimensions an analogous magnetization of localized states pinned at the zig-zag termination of graphene ribbons was observed\cite{magda2014graphene_ribbon} experimentally and studied theoretically.\cite{Yazyev_Katsnelson_PhysRevLett.100.047209,NOJIJ_PhysRevB.99.224412}

Surface and interface magnetism has already significantly impacted consumer electronics, with the giant magnetoresistance effect underpinning many data storage solutions. \cite{Fert_PhysRevLett.61.2472} Spintronics architectures incorporating magnetic boundaries promise smaller, faster, and more energy-efficient classical and quantum devices. \cite{bader2010spintronics} Understanding and exploiting the new avenue for surface and interface magnetism provided by the drumhead surface states of nodal line semimetals is thus desired. 

In this work, we investigate surface reconstruction and surface magnetism in the stoichiometric nodal loop semimetal Ca$_3$P$_2$. This material, first isolated by Xie and coworkers,\cite{ca3p2_experiment} is a form of Ca$_3$P$_2$ arranged in a hexagonal crystal structure similar to Mn$_5$Si$_3$, however according to X-ray diffraction data all Ca sites have only a partial occupancy of 90\%. Thus this arrangement yields a stable stoichiometry balancing the constituent Ca$_{2+}$ and P$_{3-}$ ionic cores. We applied a unified first principles approach, based on the SIESTA method.\cite{siestapaper} This allowed us to perform geometrical relaxation of the considered slab structures, to obtain electronic spectra for systems with partial occupancy and to extract information regarding the localization and magnetization properties of the electronic states near the Fermi energy in the bulk and on the surface of the sample.
Our results show how surface reconstruction and band bending may interfere with surface magnetism.

\section*{Results}

In this section, we detail all our findings regarding the surface states of Ca$_3$P$_2$. First, we present the structural and electronic bulk properties of the system obtained from our first principles approach. Second, we discuss structural properties and surface reconstruction of the two considered slab geometries. We then explore the spectral properties of the surface states. Finally, we turn to the discussion of surface magnetism in the stoichiometric system and beyond. 

\subsection*{Properties of the bulk}

\begin{figure}[!h]
    \centering
    \includegraphics[width=1\linewidth]{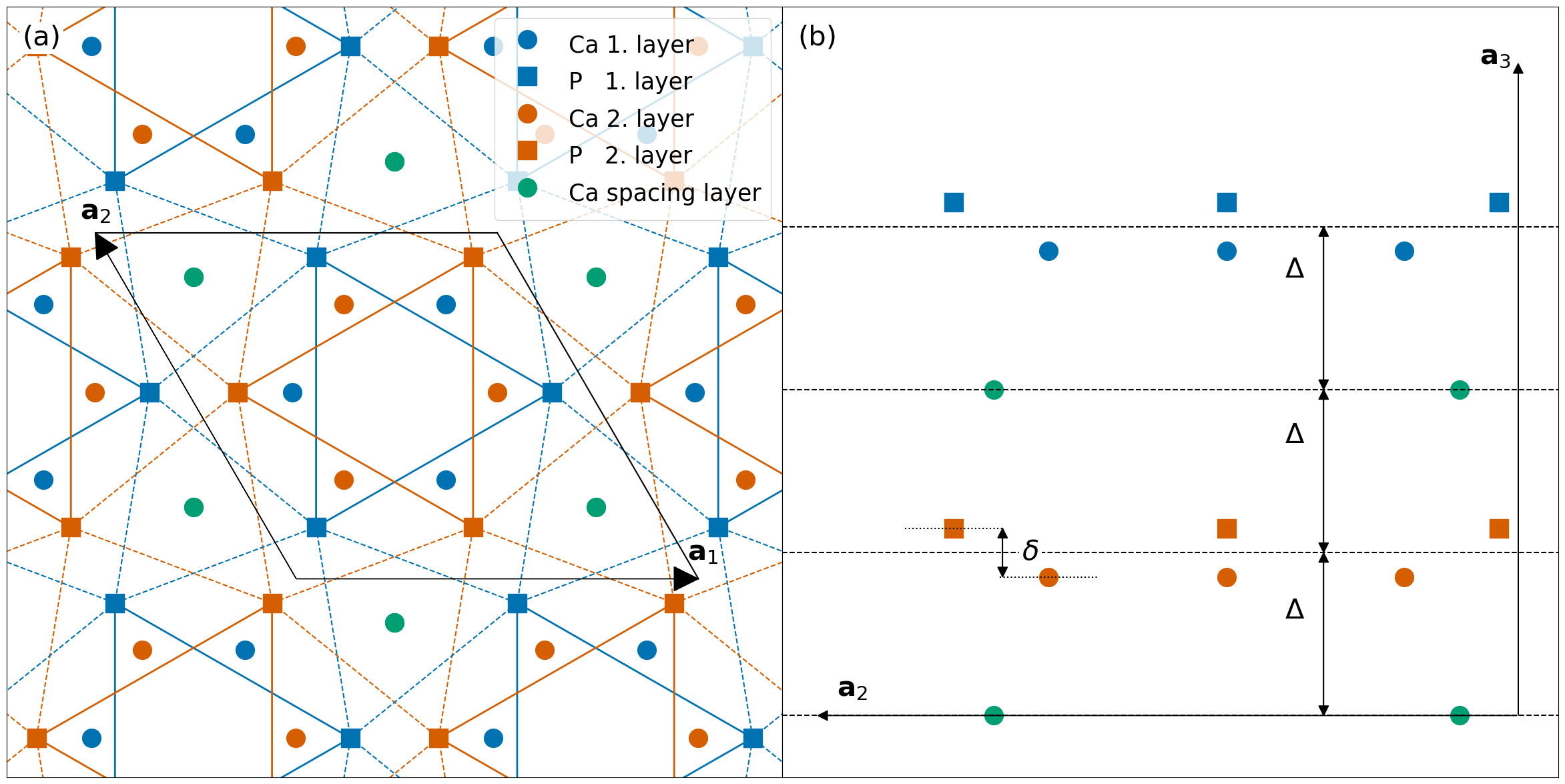}
    \caption{Real space structure of Ca$_3$P$_2$. Top (a) and side (b) view of the unit cell. Lattice vectors $\mathbf{a}_1$, $\mathbf{a}_2$ and $\mathbf{a}_3$ spanning the unit cell are denoted by arrows, Ca sites are marked by disks while P sites by squares. Blue and orange colors refer to common layer 1. and 2. respectively. Spacing layers are marked by green. }
    \label{fig:real space}
\end{figure}

The bulk crystal structure of Ca$_3$P$_2$ is depicted in Figure~\ref{fig:real space}. The bulk is spawned by lattice vectors $\mathbf{a}_1=a(1,0,0)$, $\mathbf{a}_2=a(-1/2,\sqrt{3}/2,0)$ and $\mathbf{a}_3=c(0,0,1)$. The bulk structure consists of three distinct types of layers. Two layers, denoted by blue and orange colour contain three Ca and three P sites. We shall refer to these as common layers. A third type of layer, the spacing layer, contains two Ca sites and is situated between the common layers. A unit cell has two spacing layers and two common layers. 
In a given common layer both Ca and P sites form interlocked distorted triangular lattices. This is highlighted for the P sites in Figure~\ref{fig:real space}. where all nearest neighbour P sites are connected by solid (shorter bonds) or dashed (longer bonds) lines. In the bulk, the constituent layers are equally spaced, that is $\Delta=c/4$ and the layers do not exhibit any buckling, that is. We note that in the case of slab calculations discussed below, both spacing and buckling are appreciably changed near the surface. We find that buckling is introduced in the common layer in such a way that all Ca sites are shifted together with respect to all P sites, this shift is denoted by $\delta$.
Structural relaxation of the bulk crystal yielded the parameters presented in Table~\ref{tab:bulk_lattice}. These are in general in good agreement with those reported in Ref.~\citeonline{ca3p2_experiment} although $a$ was found slightly larger and $c$ slightly lower in our calculations. 

\begin{table}[!h]
\centering
\begin{tabular}{l|llll}
Space group                           & hexagonal $P6_3/mcm$                            &                               &                            &                \\ \hline
Lattice constants (\AA)             &                                    &                               &                            &                \\ \hline
$a$                                     & 8.29415                            &                               &                            &                \\
$c$                                     & 6.70378                            &                               &                            &                \\ \hline
Atomic coordinates (fractional units) & \multicolumn{1}{l|}{site symmetry} & $\mathbf{a}_1$                & $\mathbf{a}_2$             & $\mathbf{a}_3$ \\ \hline
Ca, spacer layer                      & \multicolumn{1}{l|}{4d}            & \multicolumn{1}{l|}{$1/3$}    & \multicolumn{1}{l|}{$2/3$} & $0$            \\
Ca, common layer                       & \multicolumn{1}{l|}{6g}            & \multicolumn{1}{l|}{0.255653} & \multicolumn{1}{l|}{0}     & $1/4$          \\
P, common layer                        & \multicolumn{1}{l|}{6g}            & \multicolumn{1}{l|}{0.609573} & \multicolumn{1}{l|}{0}     & $1/4$         
\end{tabular}
\caption{\label{tab:bulk_lattice}Summary of optimized bulk lattice parameters.}
\end{table}

\begin{figure}[!h]
    \centering
    \includegraphics[width=\linewidth]{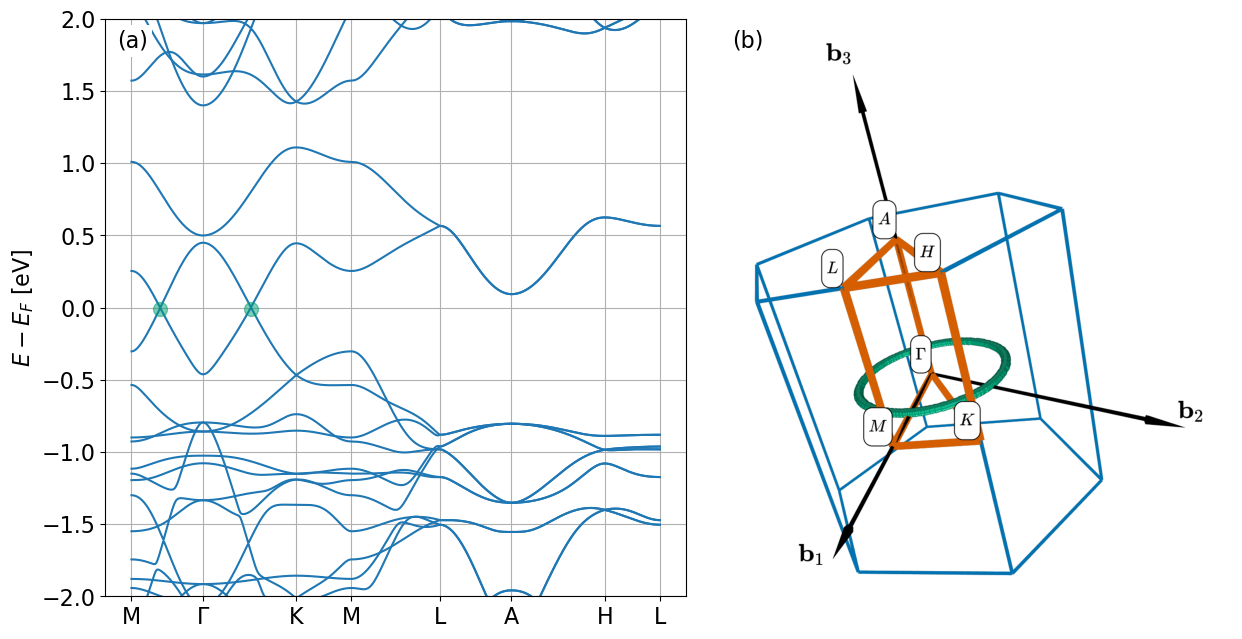}
    \caption{Electronic band structure of bulk Ca$_3$P$_2$ along high symmetry points of the three-dimensional Brillouin zone (a). Green disks marking band degeneracies in the M-$\Gamma$-K plane. Subfigure (b) depicts the Brillouin zone spawned by reciprocal lattice vectors $\mathbf{b}_1$,$\mathbf{b}_2$ and $\mathbf{b}_3$ as well as high symmetry points. The green curve, calculated as the degeneracy of the bulk bands, indicates the nodal loop.}
    \label{fig:reciprocal space}
\end{figure}

The electronic spectrum depicted in Figure~\ref{fig:reciprocal space}. shows a clear signature of nodal line semimetals, namely a continuous line of band degeneracy in the plane spawned by $\mathbf{b}_1$ and $\mathbf{b}_2$ reciprocal vectors at the Fermi level. We again note a good qualitative agreement between our results and those reported in Ref.~\citeonline{ca3p2_experiment}. However, we remark that the spectrum reported previously in Ref.~\citeonline{ca3p2theory_PhysRevB.93.205132} differs somewhat around the A-L-H plane, specifically in our case states of the conduction band approach the Fermi energy closer at the A point. This discrepancy can be attributed to the slightly different implementations of density functional theory (DFT). In our case localized pseudo atomic orbital basis was used in conjunction with a pseudopotential approach, while those calculations presented in Ref.~\citeonline{ca3p2theory_PhysRevB.93.205132} used an all electron code with plane wave basis. 

\subsection*{Slab calculations}

In the following, we present the results of our \emph{ab initio} calculations performed for slabs of a finite thickness. All presented results were obtained for slabs of 6 unit cell thickness in the $\mathbf{a}_3$ direction as our structural and magnetic calculations showed that the middle of these structures already approximated the bulk well. First, we discuss structural properties then we turn our attention to the electronic spectrum and magnetic properties. 

\subsubsection*{Structural surface reconstruction}

We considered two symmetric slab geometries differing in the type of the very first and last layer. We shall refer to slabs as capped if the first layer is a common layer, built from Ca and P atoms, while we shall call slabs terminated if the first layer is a spacing layer containing only Ca atoms.

\begin{figure}[h]
    \centering
    \includegraphics[width=0.49\linewidth]{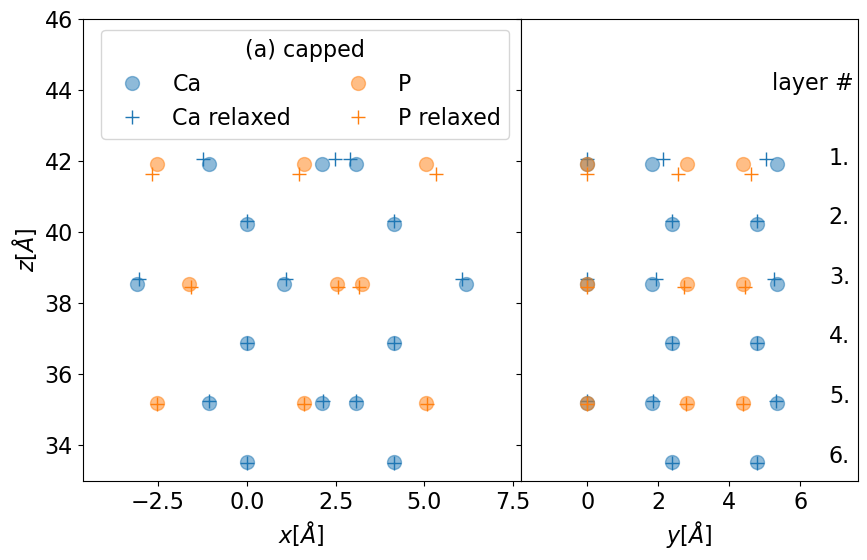}
    \includegraphics[width=0.49\linewidth]{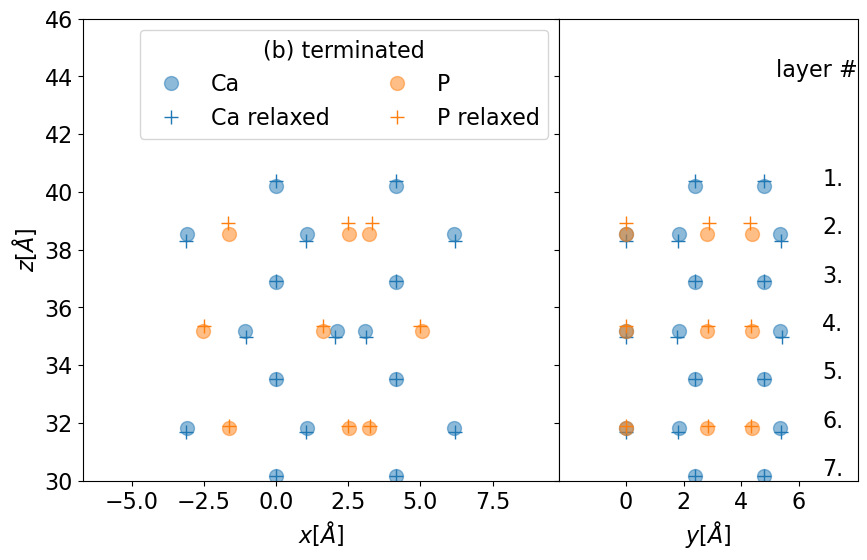}
    \caption{Structure of the surface of a slab before and after relaxation in the capped (a) and in the terminated (b) configuration.}
    \label{fig:slab_relax_geo}
\end{figure}

We performed geometry optimization on both types of slabs. The resulting arrangements are presented in Figure~\ref{fig:slab_relax_geo}. For the capped configuration depicted in subfigure (a) one can observe that P atoms move towards the bulk and Ca atoms slightly elevate out from the surface. This introduces a buckling $\delta$ in the order of $0.4$\AA~on the surface. One can observe a considerable lateral displacement as well in the case of a capped slab.
For terminated slabs shown in Figure~\ref{fig:slab_relax_geo} (b) Ca atoms, in layer 1. move slightly away from the system. On the next layer, P atoms also move away from the bulk. That is in this layer the atoms behave oppositely compared to how they did in the first common layer of the capped slab. Curiously the sample does not exhibit any substantial relaxation in the lateral direction. The observed scale of the buckling $\delta$ is similar to that seen for the capped slab. 
Surface reconstruction, as it is expected, diminishes as we move towards the bulk. We note that for capped slabs after the first unit cell, \emph{i. e.} after four layers the geometry of the bulk is recovered. In the case of terminated slabs, this recovery is somewhat slower, in particular, Ca atoms are still further than $0.1$\AA~compared to their unrelaxed position in layer 6, which is in the second unit cell.  

\subsubsection*{Electronic spectra of surface states}
For slab geometries the periodicity of the system is two-dimensional.
In this case, the high symmetry points of the bulk Brillouin zone collapse to the high symmetry points of a two-dimensional Brillouin zone. In particular, $\Gamma$ and A are mapped to $\Gamma$, L and M are mapped to M and H and K to K. In the following, we denote the wave vectors of this two-dimensional Brillouin zone with $\vec{k}$ as as opposed to the three-dimensional notation $\mathbf{k}$. 

\begin{figure}[!h]
    \centering
    \includegraphics[width=\linewidth]{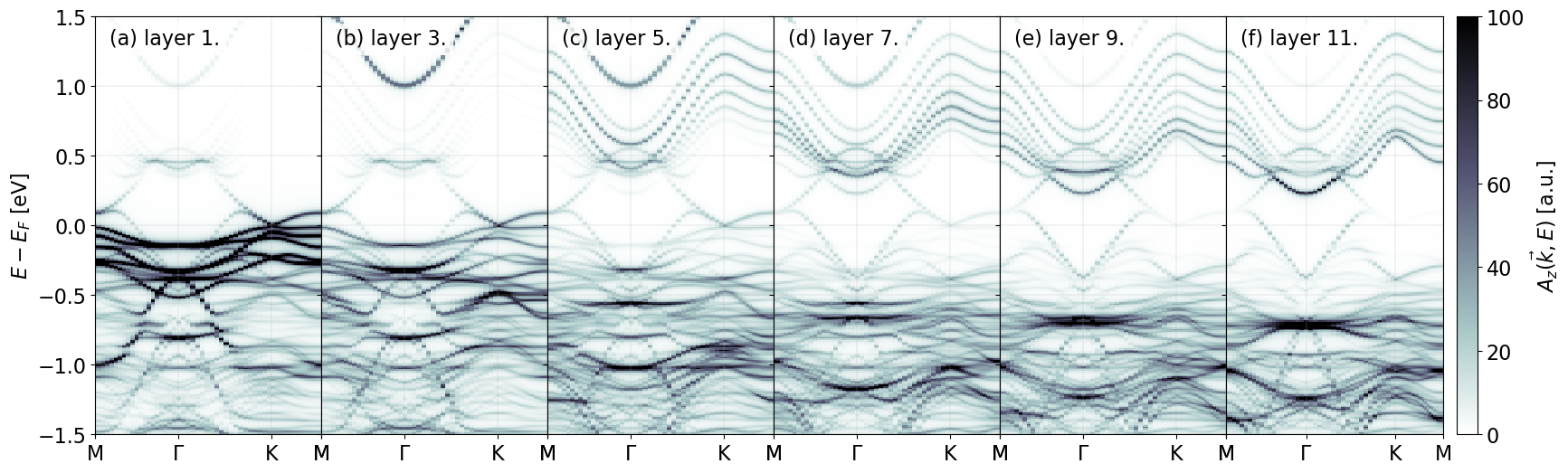}
    \caption{Layer resolved spectral function $A_z(\vec{k},E)$ for a capped slab at with stoichiometric Ca occupations.}
    \label{fig:Akz_capped}
\end{figure}

The different geometrical arrangements of the two considered slab types result in a markedly different electronic structure for the two cases. 
First, we discuss capped slabs. In Figure~\ref{fig:Akz_capped}. the layer resolved spectral function, $A_z(\vec{k},E)$ for capped slabs is shown for common layers. We note that spacing layers have a negligible spectral weight in the energy window shown thus we do not plot the spectral function localized on them.
One can observe clear signatures of band bending in the presented spectral data. The majority of the spectral weight shifts towards lower energies as one approaches the bulk. 
On the first layer states with relatively low kinetic energy appear, however, they span the whole Brillouin zone contrary to what one expects from drumhead states. These states are also just under the Fermi level. At higher energies 0.5 eV above $E_F$ on the first layer and slightly below in the layers closer to the bulk a relatively flat band also appears. 

\begin{figure}[!h]
    \centering
    \includegraphics[width=\linewidth]{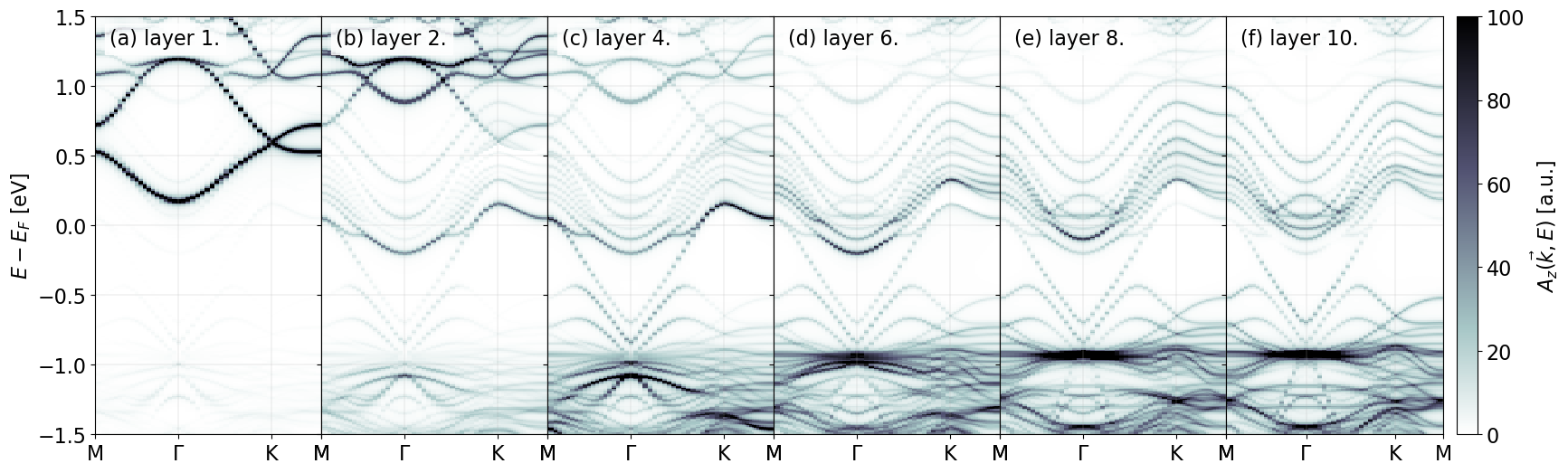}
    \caption{Layer resolved spectral function $A_z(\vec{k},E)$ for a terminated slab with stoichiometric Ca occupations.}
    \label{fig:Akz_termin}
\end{figure}

Slabs in the terminated geometry exhibit less band bending, and the electronic states are lowered in energy compared to the capped arrangement, as is evident in Figure~\ref{fig:Akz_termin}. In this geometry, the first layer is a spacing layer with a considerable spectral weight localized on it, albeit all of this weight is above the Fermi level and forms a graphene-like band structure. As it was in the case for capped slab spacing layers further inside the system have negligible spectral weight in the shown energy window. In the common layers close to the surface the spectral function is prominent in a band situated close to the Fermi level. Comparing the features of this band one notes similarities to the band structure observed at 0.5 eV above the Fermi level in the capped system. This spectral feature might be best attributed to the drumhead surface states, however as in the case of the capped slab, the flat spectral features around the Fermi energy extend further in the Brillouin zone. 

As we discussed above, both slab configurations exhibit flat spectral features in the vicinity of the Fermi level, thus a natural expectation is that interactions drastically impact the system for instance by germinating a finite magnetic moment close to the surface of the slabs. Contrary to these expectations the magnetization of the system remains rather low in both cases. The largest moments in both cases are localized on common layers close to the surface but are only on the order of $10^{-4}\mu_B$. 

\begin{figure}
    \centering
    \includegraphics[width=\linewidth]{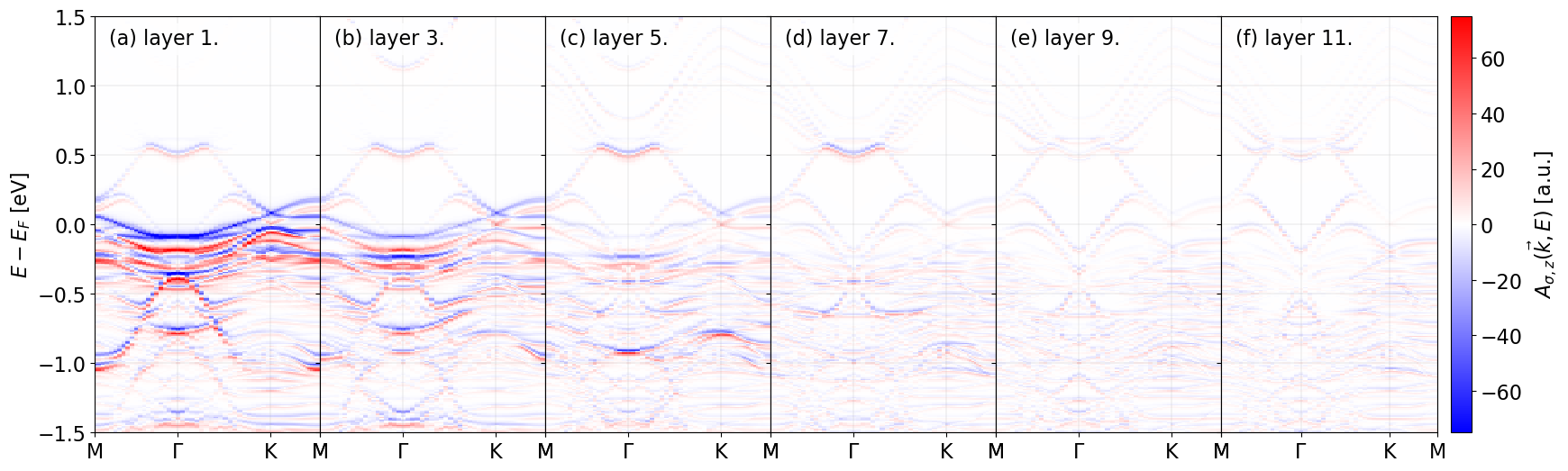}
    \caption{Layer resolved magnetic spectral function $A_{\sigma,z}(\vec{k},E)$ for a capped slab at $88.8\%$ Ca occupancy. }
    \label{fig:Aksz_capped}
\end{figure}

\begin{figure}
    \centering
    \includegraphics[width=\linewidth]{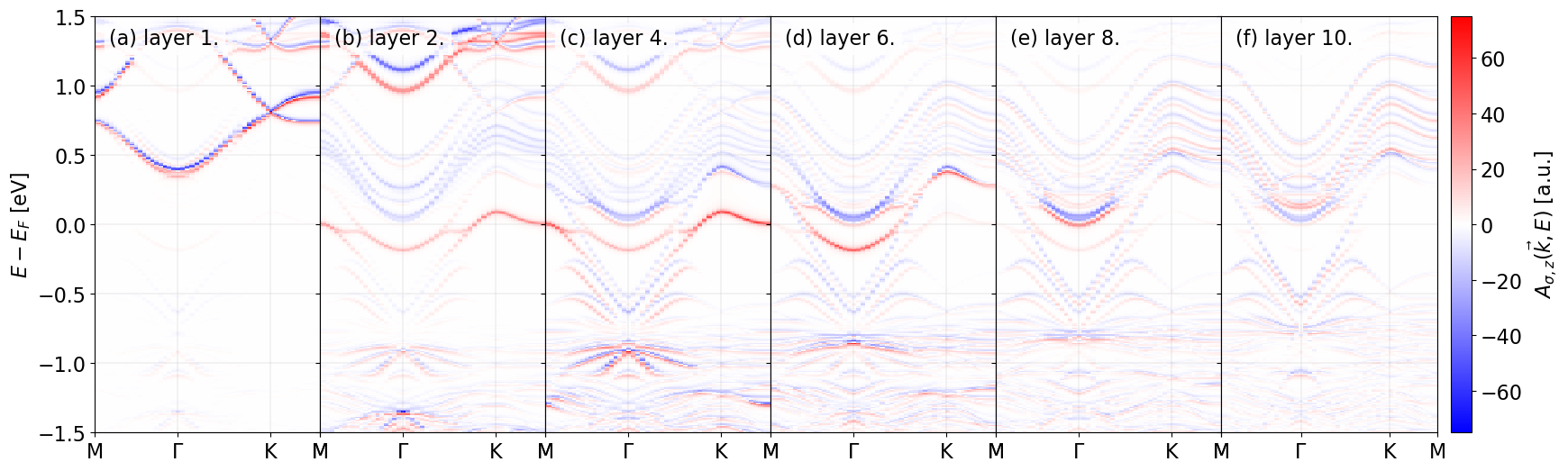}
    \caption{Layer resolved spin spectral function $A_{\sigma,z}(\vec{k},E)$ for a terminated slab at $88.8\%$ Ca occupancy. }
    \label{fig:Aksz_termin}
\end{figure}

In an experimental setting, such as Ref.~\citeonline{ca3p2_experiment} the Ca site occupancy can differ from the stoichiometric $90\%$. We now discuss magnetic spectral functions for $88.8\%$ occupancy depicted in Figure~\ref{fig:Aksz_capped}. and Figure~\ref{fig:Aksz_termin}. In these cases an appreciable magnetic moment $\approx 0.1 \mu_B$ will be localized on atoms near the surface. The magnetic spectral functions depicted in Figures~\ref{fig:Aksz_capped}. and \ref{fig:Aksz_termin}. clearly show that at this occupancy electronic states in the two spin channels are split on common layers close to the surface, while towards the bulk this splitting is reduced. We note that the total spectral function, not shown here, in this case differs only slightly compared to those at the stoichiometric occupancy.

\subsubsection*{Magnetism below the stoichiometric limit}

Above we established that at slightly diminished Ca occupancy electronic states of opposing spin channels are energetically split. Below we give further analysis of this state from the perspective of the magnetic moments localized at atomic sites close to the surface.

\begin{figure}[h]
    \centering
    \includegraphics[width=0.6\linewidth]{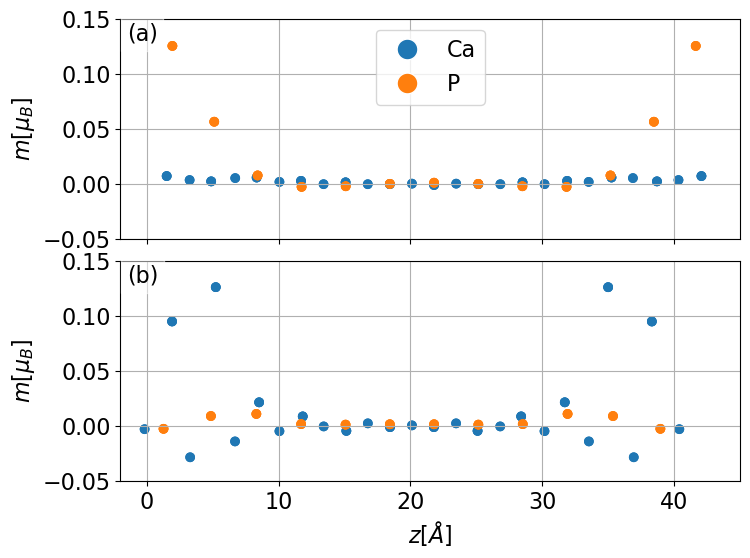}
    \caption{Layer dependent magnetization $m$ as a function of height $z$ for capped and terminated configurations $88.8\%$ Ca concentration. }
    \label{fig:m_vs_z}
\end{figure}

\begin{figure}
    \centering
    \includegraphics[width=\linewidth]{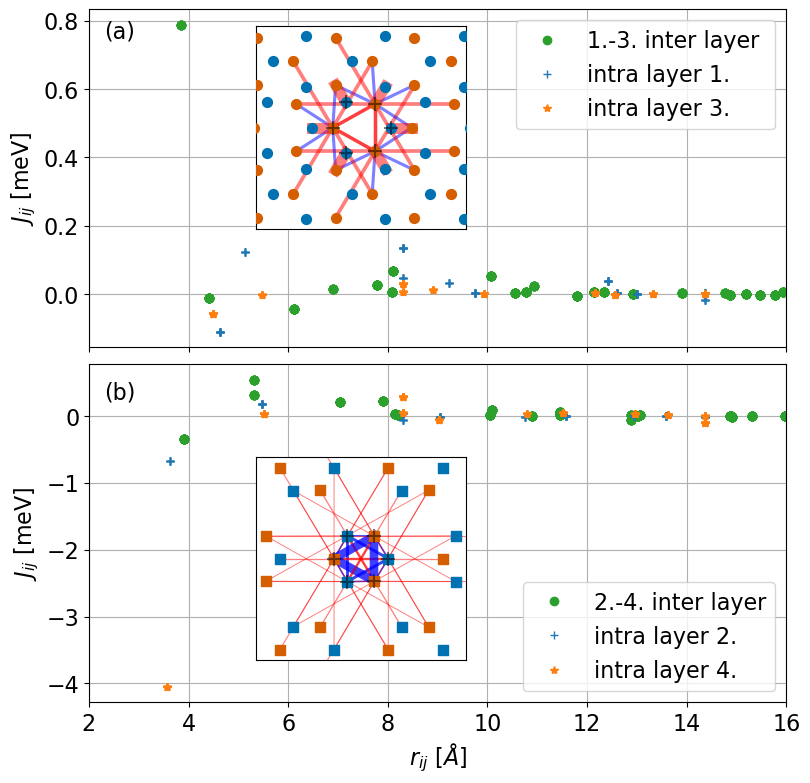}
    \caption{Effective Heisenberg couplings $J_{ij}$ between magnetized layers in the capped (a) and terminated (b) configuration for $88.8\%$ Ca concentration. In subfigure (a) the P atoms of layers 1. and 3. are denoted by orange and blue circles. In subfigure (b) the Ca atoms of layers 2. and 4. are denoted by squares. For both cases, the thickness of the connecting line is proportional to the interaction strength normalized to the strongest interaction. Red colors refer to ferromagnetic coupling while blue colors signal antiferromagnetic coupling. Black crosses mark sites inside the unit cell which were taken as the origin for the calculated exchange interactions.}
    \label{fig:Jij}
\end{figure}

In Figure~\ref{fig:m_vs_z}. magnetic moments $m$ as a function of the $z$ coordinate of atoms in the slab are presented for capped (a) and terminated (b) configurations respectively. In the capped configuration the largest magnetic moment is located on the surface, that is on layer 1., while no appreciable magnetism can be observed beyond the first unit cell. Magnetic moments on the surface for the capped configuration are strictly localized on the P atoms. 
In the case of a terminated slab, the very first layer is not magnetized. Layers 2. and 4. carry most of the magnetic moment in the system. In this case spacing layers around them also develop a small magnetic moment. Magnetic moments in the terminated configurations, as opposed to the capped configurations, are associated with the Ca atoms. As for the capped configuration, we observe no magnetization in the bulk for the terminated configuration.
The magnetic moments discussed above were obtained in a collinear reference frame calculation. By fitting the energetics of infinitesimal spin rotations to the electronic structure we extracted effective Heisenberg exchange interactions.\cite{NOJIJ_PhysRevB.99.224412} For both types of slabs we calculated interactions only between sites which exhibited a larger magnetic moment than 0.05 $\mu_B$, that is in the case of capped slabs P atoms on layers 1. and 3. and for terminated slabs Ca sites on layers 2. and 4. were taken into account. The obtained interactions are depicted in Figure~\ref{fig:Jij}. 
Again the results show marked differences between capped and terminated slabs. In the case of capped slabs, the strongest interaction is ferromagnetic between P atoms with the smallest lateral displacement but in a different layer. Inspection of further neighbour terms reveals predominantly ferromagnetic couplings. We thus conclude that for capped slabs the surface layer shows a ferromagnetic arrangement. 
A more complicated magnetic texture can be inferred in the terminated slab. Coupling between nearest neighbours on layer 4. is the strongest coupling with an antiferromagnetic nature. Similarly, antiferromagnetic coupling can be observed for nearest neighbours on layer 2. This observation suggests that in terminated slabs localized moments will not prefer a co-linear arrangement, but rather a 120$^{\circ}$ Neél configuration is expected.\cite{fazekas1999lecture}

\section*{Discussion}

In this manuscript, we report a concise \emph{ab initio} study of surface reconstruction and surface localized magnetization in the nodal loop semimetal Ca$_3$P$_2$. Our calculations go beyond previous studies\cite{ca3p2theory_PhysRevB.93.205132} for slabs of Ca$_3$P$_2$ where a tight-binding model was fitted to the bulk band structure and the obtained slab spectrum was calculated in the absence of surface reconstruction. 
We investigated two types of slabs with different surface termination. We show that in the stoichiometric limit even though states with reduced kinetic energy are localized close to the surface, the interplay of geometrical reconstruction and band bending pushes these states just below the Fermi level impeding the emergence of surface magnetism. Even though the bulk exhibited a pronounced nodal loop, in the slab calculations it is hard to identify the drumhead surface states because of other non-dispersive bands present in the spectrum.
Deviating slightly from the stoichiometric limit the surface is shown to be magnetic with a magnetic texture depending strongly on surface configuration. 
As an alternative to adjusting the occupancy of Ca sites, in an experimental setup where slabs of Ca$_3$P$_2$ are investigated, a similar emergence of magnetism could be achieved by controlling the Fermi level of the system with a gate voltage.

The results presented here underscore the significance of surface reconstruction in altering the magnetic properties of nodal loop semimetals, emphasizing that even minimal deviations in atomic arrangements can have profound effects on surface magnetism. Future research focusing on fine-tuning these surface characteristics may pave the way for practical applications where the unique properties of nodal loop semimetals could be harnessed to engineer novel magnetic and electronic functionalities.

\section*{Methods}

In all our \emph{ab initio} calculations we used the SIESTA approach to DFT.\cite{siestapaper} Bulk calculations were performed with a mesh of $21\times 21\times 21$ $k$-points while slab calculations were done on a $21\times 21$ mesh. A real space grid cutoff of $1000$ Ry was employed. 
The exchange and correlation energies were approximated by the use of the GGA-PBE functional.\cite{PBE_PhysRevLett.77.3865}
Lattice relaxation was performed with a maximum force tolerance of $0.008$ eV/\AA.
All relaxed structures were obtained by relaxing the system with fully occupied Ca sites and without spin polarization.
To treat partial occupancy of the Ca sites we employed the virtual crystal approximation (VCA) in the spirit of previous works\cite{ca3p2_experiment,ca3p2theory_PhysRevB.93.205132} as it is implemented in the SIESTA package.\cite{siesta_VCA_poloni2010efficient} All our calculations employing VCA used a mixture of Ca and Ar atoms. We justify this choice by noting that in the experiments first reporting the isolation of Ca$_3$P$_2$ samples were prepared in Ar atmosphere.\cite{ca3p2_experiment}
We note that VCA is a rough approximation compared to the more sophisticated Coherent Potential Approximation (CPA).\cite{soven_cpa_PhysRev.156.809,gyorffy_cpa_PhysRevB.5.2382} To the best of our knowledge, no existing codes combine CPA with geometry relaxation, limiting our ability to fully capture the complexities of systems with partial occupancy. Future advances integrating CPA with geometry relaxation would improve accuracy in such studies.
As the constituent elements were rather light in all of our calculations we neglected spin-orbit coupling. Spin-polarized calculations were carried out in a colinear reference frame. Magnetic moments were obtained by standard Mulliken charge analysis provided by SIESTA\cite{siestapaper}. To find the ground state spin configurations \emph{e. g.} those depicted in Figure~\ref{fig:m_vs_z}. we considered three initial spin configurations and let the electronic system converge. The three configurations considered were the ferromagnetic (all atoms polarized), the paramagnetic (no atoms polarized), and the opposing surface antiferromagnetic (common layers on the two opposing surfaces polarized in the opposite direction). In all cases, the ferromagnetic initial configuration arrived at the lowest total energy yielding a pattern reported here in Figure~\ref{fig:m_vs_z}.

The layer- and spin-resolved spectral function were calculated with the help of the \textsc{sisl} post-processing package.\cite{zerothi_sisl} For a given wave number $\vec{k}$ SIESTA yields the overlap matrix $S(\vec{k})$ and spin channel dependent Hamiltonian $H^s(\vec{k})$. Solution of the generalized eigenvalue problem $H^s(\vec{k})\varphi^s_n(\vec{k})=E^s_n(\vec{k})S(\vec{k})\varphi^s_n(\vec{k})$ in turn delivers the spectrum $E^s_n(\vec{k})$ and eigenvectors $\varphi^s_n(\vec{k})$ in the basis of pseudo atomic orbitals.
With these, the spin-resolved spectral function is given by
\begin{align*}
    A_{s,z}(\vec{k},E)=-\frac{1}{\pi}\mathrm{Im}\left (\sum_{l\in z} \frac{[\varphi^s_n(\vec{k})]^*_l \left [ S(\vec{k})\varphi^s_n(\vec{k}) \right]_l }{E+\mathrm{i}\eta-E^s_n(\vec{k})} \right),
\end{align*}
where the index $l$ runs over all orbitals of atoms associated to layer $z$, and $\eta$ is small positive number.  The total and magnetic spectral functions are simply given by $A_{z}(\vec{k},E)=A_{\uparrow,z}(\vec{k},E)+A_{\downarrow,z}(\vec{k},E)$ and $A_{\sigma,z}(\vec{k},E)=A_{\uparrow,z}(\vec{k},E)-A_{\downarrow,z}(\vec{k},E)$.

We obtained the magnetic exchange interactions by evaluating the Lichtenstein formula\cite{liechtenstein1987local} adapted to the case of non-orthogonal basis functions used by the SIESTA package.\cite{NOJIJ_PhysRevB.99.224412,Relat_NOJIJ_PhysRevB.108.214418} The isotropic Heisenberg interaction $J_{ij}$ between two atomic sites $i$ and $j$ is given by 
\begin{align*}
J_{ij}=\frac{1}{2\pi}\int_{-\infty}^{E_F}\mathrm{d}E\mathrm{Im}\mathrm{Tr}_l\left [ (H^\uparrow_{ii}-H^\downarrow_{ii}) G^\uparrow_{ij}(E) (H^\uparrow_{jj}-H^\downarrow_{jj}) G^\uparrow_{ji}(E) \right ],
\end{align*}
where $\mathrm{Tr}_l$ denotes trace over orbitals, $(H^\uparrow_{ii}-H^\downarrow_{ii})$ contains the matrix elements of the exchange field on-site $i$ and $G^s_{ij}(E)$ are the site and spin-dependent matrix elements of the Green's function in real space. These parameters allow one to build a classical spin model defined by the Hamiltonian $\mathcal{H}=-\frac{1}{2}\sum_{i\neq j}J_{ij}\mathbf{e}_i\cdot\mathbf{e}_j$, where the $\mathbf{e}_i$ are unit vectors are aligned with the magnetization of site $i$.

\bibliography{refs}

\begin{thebibliography}{10}
\urlstyle{rm}
\expandafter\ifx\csname url\endcsname\relax
  \def\url#1{\texttt{#1}}\fi
\expandafter\ifx\csname urlprefix\endcsname\relax\def\urlprefix{URL }\fi
\expandafter\ifx\csname doiprefix\endcsname\relax\def\doiprefix{DOI: }\fi
\providecommand{\bibinfo}[2]{#2}
\providecommand{\eprint}[2][]{\url{#2}}

\bibitem{NODAL_REV_yang2018symmetry}
\bibinfo{author}{Yang, S.-Y.} \emph{et~al.}
\newblock \bibinfo{journal}{\bibinfo{title}{Symmetry demanded topological nodal-line materials}}.
\newblock {\emph{\JournalTitle{Advances in Physics: X}}} \textbf{\bibinfo{volume}{3}}, \bibinfo{pages}{1414631} (\bibinfo{year}{2018}).

\bibitem{fang2015topological}
\bibinfo{author}{Fang, C.}, \bibinfo{author}{Chen, Y.}, \bibinfo{author}{Kee, H.-Y.} \& \bibinfo{author}{Fu, L.}
\newblock \bibinfo{journal}{\bibinfo{title}{Topological nodal line semimetals with and without spin-orbital coupling}}.
\newblock {\emph{\JournalTitle{Physical Review B}}} \textbf{\bibinfo{volume}{92}}, \bibinfo{pages}{081201} (\bibinfo{year}{2015}).

\bibitem{NODAL_first_PhysRevLett.115.036807}
\bibinfo{author}{Yu, R.}, \bibinfo{author}{Weng, H.}, \bibinfo{author}{Fang, Z.}, \bibinfo{author}{Dai, X.} \& \bibinfo{author}{Hu, X.}
\newblock \bibinfo{journal}{\bibinfo{title}{Topological node-line semimetal and dirac semimetal state in antiperovskite ${\mathrm{cu}}_{3}\mathrm{PdN}$}}.
\newblock {\emph{\JournalTitle{Phys. Rev. Lett.}}} \textbf{\bibinfo{volume}{115}}, \bibinfo{pages}{036807}, \doiprefix\url{10.1103/PhysRevLett.115.036807} (\bibinfo{year}{2015}).

\bibitem{matusiak2017thermoelectric}
\bibinfo{author}{Matusiak, M.}, \bibinfo{author}{Cooper, J.} \& \bibinfo{author}{Kaczorowski, D.}
\newblock \bibinfo{journal}{\bibinfo{title}{Thermoelectric quantum oscillations in zrsis}}.
\newblock {\emph{\JournalTitle{Nature communications}}} \textbf{\bibinfo{volume}{8}}, \bibinfo{pages}{15219} (\bibinfo{year}{2017}).

\bibitem{ca3p2_experiment}
\bibinfo{author}{Xie, L.~S.} \emph{et~al.}
\newblock \bibinfo{journal}{\bibinfo{title}{{A new form of Ca3P2 with a ring of Dirac nodes}}}.
\newblock {\emph{\JournalTitle{APL Materials}}} \textbf{\bibinfo{volume}{3}}, \doiprefix\url{10.1063/1.4926545} (\bibinfo{year}{2015}).
\newblock \bibinfo{note}{083602}.

\bibitem{Roy_Interacting_nodal_PhysRevB.96.041113}
\bibinfo{author}{Roy, B.}
\newblock \bibinfo{journal}{\bibinfo{title}{Interacting nodal-line semimetal: Proximity effect and spontaneous symmetry breaking}}.
\newblock {\emph{\JournalTitle{Phys. Rev. B}}} \textbf{\bibinfo{volume}{96}}, \bibinfo{pages}{041113}, \doiprefix\url{10.1103/PhysRevB.96.041113} (\bibinfo{year}{2017}).

\bibitem{RECENT_NODAL_ARPES_PhysRevB.107.045142}
\bibinfo{author}{Song, C.} \emph{et~al.}
\newblock \bibinfo{journal}{\bibinfo{title}{Observation of spin-polarized surface states in the nodal-line semimetal ${\mathrm{sntas}}_{2}$}}.
\newblock {\emph{\JournalTitle{Phys. Rev. B}}} \textbf{\bibinfo{volume}{107}}, \bibinfo{pages}{045142}, \doiprefix\url{10.1103/PhysRevB.107.045142} (\bibinfo{year}{2023}).

\bibitem{Campetella_PhysRevB.101.165437}
\bibinfo{author}{Campetella, M.} \emph{et~al.}
\newblock \bibinfo{journal}{\bibinfo{title}{Hybrid-functional electronic structure of multilayer graphene}}.
\newblock {\emph{\JournalTitle{Phys. Rev. B}}} \textbf{\bibinfo{volume}{101}}, \bibinfo{pages}{165437}, \doiprefix\url{10.1103/PhysRevB.101.165437} (\bibinfo{year}{2020}).

\bibitem{ED_paper2021exchange}
\bibinfo{author}{Muten, J.~H.}, \bibinfo{author}{Copeland, A.~J.} \& \bibinfo{author}{McCann, E.}
\newblock \bibinfo{journal}{\bibinfo{title}{Exchange interaction, disorder, and stacking faults in rhombohedral graphene multilayers}}.
\newblock {\emph{\JournalTitle{Physical Review B}}} \textbf{\bibinfo{volume}{104}}, \bibinfo{pages}{035404} (\bibinfo{year}{2021}).

\bibitem{zhou2021half}
\bibinfo{author}{Zhou, H.} \emph{et~al.}
\newblock \bibinfo{journal}{\bibinfo{title}{Half-and quarter-metals in rhombohedral trilayer graphene}}.
\newblock {\emph{\JournalTitle{Nature}}} \textbf{\bibinfo{volume}{598}}, \bibinfo{pages}{429--433} (\bibinfo{year}{2021}).

\bibitem{hagymasi2022observation}
\bibinfo{author}{Hagym{\'a}si, I.} \emph{et~al.}
\newblock \bibinfo{journal}{\bibinfo{title}{Observation of competing, correlated ground states in the flat band of rhombohedral graphite}}.
\newblock {\emph{\JournalTitle{Science Advances}}} \textbf{\bibinfo{volume}{8}}, \bibinfo{pages}{eabo6879} (\bibinfo{year}{2022}).

\bibitem{magda2014graphene_ribbon}
\bibinfo{author}{Magda, G.~Z.} \emph{et~al.}
\newblock \bibinfo{journal}{\bibinfo{title}{Room-temperature magnetic order on zigzag edges of narrow graphene nanoribbons}}.
\newblock {\emph{\JournalTitle{Nature}}} \textbf{\bibinfo{volume}{514}}, \bibinfo{pages}{608--611} (\bibinfo{year}{2014}).

\bibitem{Yazyev_Katsnelson_PhysRevLett.100.047209}
\bibinfo{author}{Yazyev, O.~V.} \& \bibinfo{author}{Katsnelson, M.~I.}
\newblock \bibinfo{journal}{\bibinfo{title}{Magnetic correlations at graphene edges: Basis for novel spintronics devices}}.
\newblock {\emph{\JournalTitle{Phys. Rev. Lett.}}} \textbf{\bibinfo{volume}{100}}, \bibinfo{pages}{047209}, \doiprefix\url{10.1103/PhysRevLett.100.047209} (\bibinfo{year}{2008}).

\bibitem{NOJIJ_PhysRevB.99.224412}
\bibinfo{author}{Oroszl\'any, L.}, \bibinfo{author}{Ferrer, J.}, \bibinfo{author}{De\'ak, A.}, \bibinfo{author}{Udvardi, L.} \& \bibinfo{author}{Szunyogh, L.}
\newblock \bibinfo{journal}{\bibinfo{title}{Exchange interactions from a nonorthogonal basis set: From bulk ferromagnets to the magnetism in low-dimensional graphene systems}}.
\newblock {\emph{\JournalTitle{Phys. Rev. B}}} \textbf{\bibinfo{volume}{99}}, \bibinfo{pages}{224412}, \doiprefix\url{10.1103/PhysRevB.99.224412} (\bibinfo{year}{2019}).

\bibitem{Fert_PhysRevLett.61.2472}
\bibinfo{author}{Baibich, M.~N.} \emph{et~al.}
\newblock \bibinfo{journal}{\bibinfo{title}{Giant magnetoresistance of (001)fe/(001)cr magnetic superlattices}}.
\newblock {\emph{\JournalTitle{Phys. Rev. Lett.}}} \textbf{\bibinfo{volume}{61}}, \bibinfo{pages}{2472--2475}, \doiprefix\url{10.1103/PhysRevLett.61.2472} (\bibinfo{year}{1988}).

\bibitem{bader2010spintronics}
\bibinfo{author}{Bader, S.~D.} \& \bibinfo{author}{Parkin, S.}
\newblock \bibinfo{journal}{\bibinfo{title}{Spintronics}}.
\newblock {\emph{\JournalTitle{Annu. Rev. Condens. Matter Phys.}}} \textbf{\bibinfo{volume}{1}}, \bibinfo{pages}{71--88} (\bibinfo{year}{2010}).

\bibitem{siestapaper}
\bibinfo{author}{Soler, J.~M.} \emph{et~al.}
\newblock \bibinfo{journal}{\bibinfo{title}{The siesta method for ab initio order-n materials simulation}}.
\newblock {\emph{\JournalTitle{Journal of Physics: Condensed Matter}}} \textbf{\bibinfo{volume}{14}}, \bibinfo{pages}{2745}, \doiprefix\url{10.1088/0953-8984/14/11/302} (\bibinfo{year}{2002}).

\bibitem{ca3p2theory_PhysRevB.93.205132}
\bibinfo{author}{Chan, Y.-H.}, \bibinfo{author}{Chiu, C.-K.}, \bibinfo{author}{Chou, M.~Y.} \& \bibinfo{author}{Schnyder, A.~P.}
\newblock \bibinfo{journal}{\bibinfo{title}{${\mathrm{ca}}_{3}{\mathrm{p}}_{2}$ and other topological semimetals with line nodes and drumhead surface states}}.
\newblock {\emph{\JournalTitle{Phys. Rev. B}}} \textbf{\bibinfo{volume}{93}}, \bibinfo{pages}{205132}, \doiprefix\url{10.1103/PhysRevB.93.205132} (\bibinfo{year}{2016}).

\bibitem{fazekas1999lecture}
\bibinfo{author}{Fazekas, P.}
\newblock \emph{\bibinfo{title}{Lecture notes on electron correlation and magnetism}}, vol.~\bibinfo{volume}{5} (\bibinfo{publisher}{World scientific}, \bibinfo{year}{1999}).

\bibitem{PBE_PhysRevLett.77.3865}
\bibinfo{author}{Perdew, J.~P.}, \bibinfo{author}{Burke, K.} \& \bibinfo{author}{Ernzerhof, M.}
\newblock \bibinfo{journal}{\bibinfo{title}{Generalized gradient approximation made simple}}.
\newblock {\emph{\JournalTitle{Phys. Rev. Lett.}}} \textbf{\bibinfo{volume}{77}}, \bibinfo{pages}{3865--3868}, \doiprefix\url{10.1103/PhysRevLett.77.3865} (\bibinfo{year}{1996}).

\bibitem{siesta_VCA_poloni2010efficient}
\bibinfo{author}{Poloni, R.}, \bibinfo{author}{Iniguez, J.}, \bibinfo{author}{Garcia, A.} \& \bibinfo{author}{Canadell, E.}
\newblock \bibinfo{journal}{\bibinfo{title}{An efficient computational method for use in structural studies of crystals with substitutional disorder}}.
\newblock {\emph{\JournalTitle{Journal of Physics: Condensed Matter}}} \textbf{\bibinfo{volume}{22}}, \bibinfo{pages}{415401} (\bibinfo{year}{2010}).

\bibitem{soven_cpa_PhysRev.156.809}
\bibinfo{author}{Soven, P.}
\newblock \bibinfo{journal}{\bibinfo{title}{Coherent-potential model of substitutional disordered alloys}}.
\newblock {\emph{\JournalTitle{Phys. Rev.}}} \textbf{\bibinfo{volume}{156}}, \bibinfo{pages}{809--813}, \doiprefix\url{10.1103/PhysRev.156.809} (\bibinfo{year}{1967}).

\bibitem{gyorffy_cpa_PhysRevB.5.2382}
\bibinfo{author}{Gyorffy, B.~L.}
\newblock \bibinfo{journal}{\bibinfo{title}{Coherent-potential approximation for a nonoverlapping-muffin-tin-potential model of random substitutional alloys}}.
\newblock {\emph{\JournalTitle{Phys. Rev. B}}} \textbf{\bibinfo{volume}{5}}, \bibinfo{pages}{2382--2384}, \doiprefix\url{10.1103/PhysRevB.5.2382} (\bibinfo{year}{1972}).

\bibitem{zerothi_sisl}
\bibinfo{author}{Papior, N.}
\newblock \bibinfo{title}{sisl: v0.14.3}, \doiprefix\url{10.5281/zenodo.597181} (\bibinfo{year}{2024}).

\bibitem{liechtenstein1987local}
\bibinfo{author}{Liechtenstein, A.~I.}, \bibinfo{author}{Katsnelson, M.}, \bibinfo{author}{Antropov, V.} \& \bibinfo{author}{Gubanov, V.}
\newblock \bibinfo{journal}{\bibinfo{title}{Local spin density functional approach to the theory of exchange interactions in ferromagnetic metals and alloys}}.
\newblock {\emph{\JournalTitle{Journal of Magnetism and Magnetic Materials}}} \textbf{\bibinfo{volume}{67}}, \bibinfo{pages}{65--74} (\bibinfo{year}{1987}).

\bibitem{Relat_NOJIJ_PhysRevB.108.214418}
\bibinfo{author}{Mart\'{\i}nez-Carracedo, G.} \emph{et~al.}
\newblock \bibinfo{journal}{\bibinfo{title}{Relativistic magnetic interactions from nonorthogonal basis sets}}.
\newblock {\emph{\JournalTitle{Phys. Rev. B}}} \textbf{\bibinfo{volume}{108}}, \bibinfo{pages}{214418}, \doiprefix\url{10.1103/PhysRevB.108.214418} (\bibinfo{year}{2023}).

\end{thebibliography}

\section*{Acknowledgements (not compulsory)}

The authors wish to express their gratitude to Jaime Ferrer, Gabriel Martinez Carracedo, László Szunyogh and László Udvardi, for valuable discussions and their comments regarding the present work.

This research was supported by the Ministry of Culture and Innovation and the National Research, Development and Innovation Office within the Quantum Information National Laboratory of Hungary (Grant No. 2022-2.1.1-NL-2022-00004) and by NKFIH Grants No. K131938, K134437 and K142179.  A.A. greatly acknowledges the support from Stipendium Hungaricum No. 249316. 

\section*{Author contributions statement}

A.~A. performed first-principles calculations and prepared figures. A.~G.~F. provided pseudo potentials and with J.~K. analysed the data. O.~L. conceived the project. All authors contributed equally to the writing and revision of the manuscript.

\section*{Additional information}

To include, in this order: \textbf{Accession codes} (where applicable); \textbf{Competing interests} (mandatory statement). 
The corresponding author is responsible for submitting a \href{http://www.nature.com/srep/policies/index.html#competing}{competing interests statement} on behalf of all authors of the paper. This statement must be included in the submitted article file.

\end{document}